\documentclass[twocolumn,showpacs,preprintnumbers,pra,amsmath,amssymb,amsfonts,amsthm]{revtex4}
\usepackage{amssymb}
\usepackage{amsbsy}
\usepackage{epsf}

\usepackage[cp866nav]{inputenc}
\usepackage{revsymb}
\usepackage[T2A,OT1]{fontenc}
\usepackage{graphicx}

\begin{document}

\title
{Enhancement of coherence in qubits due to interaction with environment}
\author{V.V. Ignatyuk$^{a)}$, V.G. Morozov$^{b)}$}
\affiliation{a) Institute for Condensed Matter Physics, 1
Svientsitskii Street, 79011, Lviv, Ukraine\\
b) Moscow State Technical University of Radioengineering,
Electronics, and Automation, \nolinebreak[0]  Vernadsky
Prospect 78, 119454 Moscow, Russia}

\date{\today}

\begin{abstract}

The influence of the initial preparation on dephasing in open quantum dynamics
is studied using an exactly solvable model of a two-level system (qubit) interacting with
a bosonic bath. It is found that for some classes of non-selective preparation measurements,
 qubit-bath correlations lead to a significant  enhancement of coherence in the qubit at the
 initial stage of evolution. The time behavior of the qubit purity and entropy in the regime
   of enhancement of coherence is considered for different temperatures and coupling
    strengths.

\end{abstract}

\maketitle

\section{Introduction\label{Introduct}}
 An important point in the dynamics of open quantum systems is the initial state preparation.
Due to unavoidable interactions, there generally exist initial correlations between a system and its environment. Thus any physical process of preparation of the initial state of the system will affect the state of the environment as well.
 The question then arises of how the preparation procedure and initial correlations influence the subsequent evolution of the open system. Different aspects of this question were discussed by  many authors (see, e.g.,
 Refs.~\cite{PRA2010,PRA2012,PRA2013,MMR2012,Luczka} and references therein). Of special interest is the decoherence phenomenon
(the environmentally induced destruction of quantum coherence).
For instance, decoherence plays a crucial role in the dynamics of
two-state systems (qubits) which are the elementary carries of
quantum information~\cite{q-measurement1,NL2012,39min}.

At first sight it is natural to expect that  initial correlations
between an open quantum system and an environment with a huge (or
even infinite) number of degrees of freedom would increase the
decoherence rate. This is indeed the  case for some preparation
procedures~\cite{MMR2012,PRA2013,PRA2013-2} but is not
true in general. We refer to the paper~\cite{Luczka}, where the
authors demonstrated with
 examples of specific qubit-environment models that for some
initial system-environment correlated states, the ``purity'' of the qubit is
greater than in the case of initially uncorrelated states.
 Although the results of Ref.~\cite{Luczka} rely on a somewhat artificial
 assumption that the environment is initially prepared in a pure quantum state,
 the fact that quantum coherence in open systems can be enhanced due to
  system-environment correlations seems as itself to deserve thorough studies.
   It could  provide, for instance, a way for constructing dynamics of qubit registers with interesting and very promising properties.

In this paper we aim to examine in detail the connection between
 the initial preparation and the qubit dynamics. In particular, this allows us
 to establish the conditions for appreciable environment-induced enhancement of quantum
  coherence in a qubit. A somewhat unexpected result is that
   the purity of a qubit state can even increase with time.
  The principal difference between our analysis and that
   presented in Ref.~\cite{Luczka} lies in the interpretation of
    initial states of the composite  (qubit plus environment) system.
As we have already mentioned, in Ref.~\cite{Luczka} the initial states  were
taken in the  form which could illustrate the role of qubit-environment correlations but, unfortunately, these states are very unlikely to be physically realizable.
  In the present paper we consider more realistic preparation procedures
  based on quantum measurements~\cite{BP-Book,q-measurement}.

  The structure of the paper is as follows. In Sec.~\ref{secII} we give a brief review of
  measurement schemes for open quantum systems and applications to qubits.
  In Sec.~\ref{secIII} we treat time evolution of a qubit coupled to a bosonic environment
  through a dephasing interaction. Exact expressions are given for the time-dependent elements
   of the qubit density matrix when the initial state is prepared by a selective
   or non-selective measurement. Our central goal in this section is to
    demonstrate that for a large class of initial states prepared by non-selective
    measurements, the coherences (off-diagonal elements of the qubit density matrix)
    increase with time at the initial stage of evolution. In Sec.~\ref{secIV},
  we consider the time behavior of the qubit purity and entropy in the regime
   of enhancement of coherence. Conclusions are drawn in Sec.~\ref{secV}.

\section{Selective and non-selective preparation measurements\label{secII}}

As an introduction to our subsequent development, we start with a brief
 discussion of  a rather general quantum measurement scheme
which can be used to construct statistical ensembles describing
initial states of real open systems. We then apply this scheme to
 a qubit interacting with its environment.

Suppose that at all times $0 < t$ an open system $S$ is in thermal
equilibrium with a heat bath $B$, and at time zero one makes a
measurement on the system $S$ only. According to general
principles of quantum measurement theory
\cite{BP-Book,q-measurement,Krauss}, the state of the composite
system ($S+B $) after the measurement is described by the
density matrix
 \begin{equation}
    \label{rho-SB-0}
 \varrho^{}_{SB}(0)= \sum_{m}\Omega^{}_{m}\varrho^{}_{\text{eq}}\Omega^{\dagger}_{m},
 \end{equation}
where $\varrho^{}_{\text{eq}}$ is the equilibrium density matrix at temperature $T$. Operators $\Omega^{}_{m}$ act in the Hilbert space of the system $S$ and correspond to
 possible outcomes $m$ of the measurement.
In a particular case of a
 \textit{selective measurement\/}, the system $S$ is prepared in some
pure state $|\psi\rangle $. Then the sum in Eq. (\ref{rho-SB-0}) collapses into a
single term, so that
  \begin{equation}
    \label{rho-SB-selec}
     \varrho^{}_{SB}(0)=
     \frac{1}{Z}\, P^{}_{\psi}\varrho^{}_{\text{eq}}P^{}_{\psi},
  \end{equation}
where $ P^{}_{\psi}=|\psi\rangle \langle\psi| $ is the projector onto the quantum state $|\psi\rangle $ and $Z$ is the normalization factor. In general, the density matrix (\ref{rho-SB-0}) describes the resulting
ensemble after a \textit{non-selective measurement\/} in which
the outcome $m$  may be viewed as a classical random number with the
probability distribution
    \begin{equation}
   \label{p-m}
  w(m)=\text{Tr}\left\{F^{}_{m}\varrho^{}_{\rm eq}\right\}.
  \end{equation}
Positive operators
  $F^{}_{m}=\Omega^{\dagger}_{m}\Omega^{}_{m}$ are called the ``effects''.
   Here and in what follows, $\text{Tr}$ denotes the trace over the Hilbert space of the
    composite ($S+B $) system, while the symbols $\text{Tr}^{}_{S}$ and $\text{Tr}^{}_{B}$
     will be used to denote the partial traces over the Hilbert spaces of the system
     $S$ and the heat bath, respectively.
 In order that $w(m)$  be normalized to 1, the effects $F^{}_{m}$ must satisfy the
 normalization condition (the resolution of the identity)
    \begin{equation}
    \label{Fm-NormCond}
  \sum_{m}F^{}_{m}\equiv \sum_{m}\Omega^{\dagger}_{m}\Omega^{}_{m}=I,
  \end{equation}
where $I$ is the unit operator.

 The precise form of $\Omega^{}_{m}$ is determined by the details of the measuring device.
 We restrict ourselves to physical situations in which some observable $A$ with a discrete,
 non-degenerate spectrum $A^{}_{m}$ is measured. Then we have~\cite{BP-Book,q-measurement}
  \begin{equation}
    \label{Omega-sqrt}
  \Omega^{}_{m}=U^{}_{m}F^{1/2}_{m},
  \end{equation}
 where $F^{1/2}_{m}$ is the square root of $F^{}_{m}$, and
 the unitary operator $U^{}_{m}$ describes the disturbance of the system $S$ by the measurement device. Formula (\ref{Omega-sqrt}) is applicable  even to
 \textit{approximate\/} measurements  where the spectrum $A^{}_{m}$ is measured with
  finite resolution~\cite{BP-Book,q-measurement}.
We will be content, however, with performing the analysis for the
case of
 infinite resolution when the effects are written as
  \begin{equation}
    \label{Fm-ideal}
   F^{}_{m}= |\psi^{}_{m}\rangle \langle\psi^{}_{m}|
  \end{equation}
  and Eq.~(\ref{Omega-sqrt}) yields
  \begin{equation}
     \label{Omega-m-gen}
  \Omega^{}_{m}=|\varphi^{}_{m}\rangle \langle \psi^{}_{m}|
  \end{equation}
  with
  \begin{equation}
    \label{phi-m}
  |\varphi^{}_{m}\rangle=U^{}_{m} |\psi^{}_{m}\rangle\, .
  \end{equation}
 A few remarks are needed here. Whereas the states $|\psi^{}_{m}\rangle$ form an
   orthonormal basis, for the transformed states  $|\varphi^{}_{m}\rangle$
   this is true only if the unitary operators $U^{}_{m}$  are identical for all
  outcomes $m$ (i.e., $U^{}_{m}=U$).
    In such cases, since  $\Omega^{}_{m}\Omega^{\dagger}_{m}=|\varphi^{}_{m}\rangle\langle\varphi^{}_{m}|$, we
     have, in addition to Eq.~(\ref{Fm-NormCond}),  another resolution of the identity
      \begin{equation}
        \label{Phi-norm-cond}
      \sum_{m}\Omega^{}_{m}\Omega^{\dagger}_{m}=I\,.
      \end{equation}
 In other words, to the  measurement scheme defined in terms of the effects
  (\ref{Fm-ideal}) and the $\Omega$-operators (\ref{Omega-m-gen}) there corresponds the
  ``dual scheme'' characterized by
   \begin{equation}
     \label{DualScheme}
     \begin{array}{l}
    \widetilde{F}^{}_{m}=UF^{}_{m}U^{\dagger}\equiv
    |\varphi^{}_{m}\rangle \langle \varphi^{}_{m}|\, ,
    \\[5pt]
    \widetilde{\Omega}^{}_{m}=\Omega^{\dagger}_{m}\equiv
     |\psi^{}_{m}\rangle \langle \varphi^{}_{m}|\, .
     \end{array}
   \end{equation}
 If the form of $U^{}_{m}$ depends on the outcome $m$, then
 the transformed states $|\varphi^{}_{m}\rangle$
 are not orthogonal in general. Some of these states may even be identical.

 Let us now apply the above general construction to a qubit.
 In the formal ``spin'' representation,
 the canonical orthonormal basis states of a qubit are
  \begin{equation}
    \label{qbit-can-bas}
  |0\rangle= \left( \begin{array}{c} 0\\ 1 \end{array}\right),
  \quad
  |1\rangle= \left(\begin{array}{c} 1\\ 0 \end{array}\right).
  \end{equation}
 It is well known that all pure states
  $|\psi(\vec{a})\rangle\equiv |\vec{a}\rangle $ correspond to points of the sphere
   $|\vec{a}|$ = 1, where $\vec{a}=\left(a^{}_{1},a^{}_{2},a^{}_{3}\right)\in \mathbb{R}^{3}$.
  Normalized state vectors are given by (see, e.g.,~\cite{Holevo2001})
    \begin{equation}
      \label{a-qubit}
     |\vec{a}\rangle=
     \left(
     \begin{array}{c}
     \displaystyle {e}^{-i\phi_a/2}\,\cos(\theta_a/2)\,\\[7pt]
       \displaystyle {e}^{i\phi_a/2}\,\sin (\theta_a/2)
      \end{array}
     \right),
    \end{equation}
     where $\phi_a$ and $\theta_a$ are the Euler angles of the unit vector $\vec{a}$ describing the direction of the ``spin''. They satisfy
    $a^{}_{1}+ia^{}_{2}= \sin\theta_a\, {e}^{i\phi_a}$, $a^{}_{3}= \cos\theta_a$.
     The Euler angles corresponding to the state $|-\vec{a}\rangle$ are
      \begin{equation}
       \label{Euler;-a}
       \theta^{}_{-a}=\pi - \theta^{}_{a}, \quad \phi^{}_{-a}=\phi^{}_{a} +\pi.
      \end{equation}
Using these relations together with Eq.~(\ref{a-qubit}) gives
      \begin{equation}
      \label{-a-qubit}
     |-\vec{a}\rangle=
     \left(
     \begin{array}{c}
     \displaystyle -i {e}^{-i\phi_a/2}\,\sin(\theta_a/2)\,\\[7pt]
     \displaystyle i{e}^{i\phi_a/2}\,\cos(\theta_a/2)
      \end{array}
     \right).
    \end{equation}

    Linear operators in the qubit's Hilbert space can be represented
as linear combinations of the unity operator and the Pauli
matrices. For example,  the operator
      \begin{equation}
        \label{sigma-a}
       \sigma(\vec{a})=\sigma^{}_{1}a^{}_{1} + \sigma^{}_{2}a^{}_{2}+\sigma^{}_{3}a^{}_{3}
      \end{equation}
  describes the spin component in the direction $\vec{a}$. The state vectors
  $|\pm\vec{a}\rangle$ correspond to the eigenvalues $\pm 1$ of $\sigma(\vec{a})$
   and form an orthonormal basis, i.e.,
     \begin{equation}
       \label{a-a-bas}
     |\vec{a}\rangle \langle\vec{a}|+  |-\vec{a}\rangle \langle-\vec{a}|=I.
     \end{equation}
     Note that the states $|\vec{a}\rangle$ and  $|-\vec{a}\rangle$
     are related to the canonical basis states (\ref{qbit-can-bas}) by
       \begin{equation}
         \label{a-Ua}
        |\vec{a}\rangle=U(\vec{a})|1\rangle,
        \qquad
          |-\vec{a}\rangle=U(\vec{a})|0\rangle
       \end{equation}
    with the unitary operator
      \begin{equation}
         \label{Ua}
      U(\vec{a})=\left(
       \begin{array}{cc}
          {e}^{-i\phi_a/2}\,\cos({\theta_a}/{2}) \ &
          -i{e}^{-i\phi_a/2}\,\sin({\theta_a}/{2})
          \\[7pt]
       {e}^{i\phi_a/2}\,\sin({\theta_a}/{2})\ &
       i {e}^{i\phi_a/2}\,\cos({\theta_a}{2})
      \end{array}
      \right).
      \end{equation}

 Selective measurements (\ref{rho-SB-selec}) on a qubit can be characterized by the
    projectors $P(\vec{a})= |\vec{a}\rangle \langle\vec{a}|$ while the general non-selective measurement scheme (\ref{Fm-ideal})\,--\,(\ref{phi-m}) is associated with three states: $|\vec{a}\rangle$, $|\vec{b}^{}_{1}\rangle$,
    and $|\vec{b}^{}_{2}\rangle$. The effects and the $\Omega$-operators are defined as
             \begin{equation}
           \label{QB-b12-FOmega}
           \begin{array}{ll}
           F^{}_{1}= |\vec{a}\rangle \langle\vec{a}|,
           \qquad
         & F^{}_{2}= |-\vec{a}\rangle \langle -\vec{a}|\, ,
         \\[5pt]
       \Omega^{}_{1}=  |\vec{b}^{}_{1}\rangle \langle\vec{a}|,
       \qquad
         &  \Omega^{}_{2}=  |\vec{b}^{}_{2}\rangle \langle-\vec{a}|\, .
          \end{array}
       \end{equation}
 According to Eqs.~(\ref{a-Ua}), we have
       \begin{equation}
          \label{b1b2a-rel}
        |\vec{b}^{}_{1}\rangle= U(\vec{b}^{}_{1},\vec{a}) |\vec{a}\rangle,
        \qquad
         |\vec{b}^{}_{2}\rangle= U(-\vec{b}^{}_{2},\vec{a}) |-\vec{a}\rangle,
       \end{equation}
     where the unitary operator $U(\vec{b},\vec{a})$ is expressed in terms of
     the operators (\ref{Ua}):
        \begin{equation}
           \label{Uba}
        U(\vec{b},\vec{a})= U(\vec{b})\, U^{\dagger}(\vec{a}).
        \end{equation}

Let us briefly consider some important special cases of the measurement scheme
 (\ref{QB-b12-FOmega}):

 i)  $\vec{b}^{}_{1}= \vec{a},\ \vec{b}^{}_{2}=- \vec{a}$. This is the
  simplest scheme corresponding to
  $U^{}_{m}=I$ in the general formula (\ref{phi-m}). Physically, here we are dealing
   with non-selective measurements where the measuring  device does not disturb the
    basis states $|\vec{a}\rangle$ and $|-\vec{a}\rangle$. In this case the
     $\Omega$-operators coincide with the effects:
       \begin{equation}
       \label{Uunity}
    \Omega_1=F_1=|\vec a\rangle\langle\vec a|, \qquad
    \Omega_2=F_2=|-\vec a\rangle\langle-\vec a|\, .
      \end{equation}
Clearly, the same operators correspond to the dual measurement scheme (\ref{DualScheme}).

ii) $\vec{b}^{}_{1}=\vec{b}$,\ $\vec{b}^{}_{2}=- \vec{b}$, where $\vec{b}$ is an
 arbitrary unit vector. This case corresponds to $U^{}_{m}=U\equiv U(\vec{b},\vec{a})$ in
 the formula~(\ref{phi-m}). It follows from Eqs.~(\ref{QB-b12-FOmega}) that
      \begin{equation}
      \label{QB-b-FOmega}
           \begin{array}{ll}
             F^{}_{1}= |\vec{a}\rangle \langle\vec{a}|,
             \qquad
           & F^{}_{2}= |-\vec{a}\rangle \langle -\vec{a}|\, ,
         \\[5pt]
           \Omega^{}_{1}=  |\vec{b}\rangle \langle\vec{a}|,
            \qquad
          & \Omega^{}_{2}=  |-\vec{b}\rangle \langle-\vec{a}|\, .
          \end{array}
       \end{equation}
There exists the  dual measurement scheme (\ref{DualScheme}) with
  \begin{equation}
     \label{QB-Non-norm}
   \begin{array}{ll}
     \widetilde{F}^{}_{1}= |\vec{b}\rangle \langle\vec{b}|,
     \qquad
     & \widetilde{F}^{}_{2}= |-\vec{b}\rangle \langle -\vec{b}|\, ,
     \\[5pt]
     \widetilde{\Omega}^{}_{1}=  |\vec{a}\rangle \langle\vec{b}|,
       \qquad
      & \widetilde{\Omega}^{}_{2}=  |-\vec{a}\rangle \langle-\vec{b}|\, .
   \end{array}
  \end{equation}

iii)  $\vec{b}^{}_{1}=\vec{b}$,\ $\vec{b}^{}_{2}=\vec{b}$ with an
 arbitrary unit vector $\vec{b}$. This is an example of a
non-selective measurement scheme for a qubit, which is described
by Eqs.~(\ref{Fm-ideal})-(\ref{phi-m}) with different unitary
operators $U_m$. Recalling Eqs.~(\ref{QB-b12-FOmega}), we write
    \begin{equation}
     \label{Omega-non}
       \begin{array}{ll}
       F^{}_{1}= |\vec{a}\rangle \langle\vec{a}|,
       \qquad
       & F^{}_{2}= |-\vec{a}\rangle \langle -\vec{a}|\, ,
       \\[5pt]
       \Omega^{}_{1}=  |\vec{b}\rangle \langle\vec{a}|,
       \qquad
       & \Omega^{}_{2}=  |\vec{b}\rangle \langle-\vec{a}|\, .
       \end{array}
  \end{equation}
It is easily verified that
   \begin{equation}
   \label{U-non}
   |\vec b\rangle=U_1|\vec a\rangle,\qquad |\vec b\rangle=U_2|-\vec a\rangle,
   \end{equation}
    where $U_1\ne U_2$ and are given  by
    \begin{equation}
    \label{U1U2}
     U_1=U(\vec b,\vec a),\qquad U_2=U(\vec b,-\vec a).
     \end{equation}
Note that in this case there is no dual measurement scheme.

We close this section with a remark about the above-discussed
 non-selective measurement schemes. Mathematically, all the schemes
  are generated by \textit{orthogonal\/} resolutions of the identity
   (\ref{Fm-NormCond}) where the effects $F^{}_{m}$ are projectors ($F^{2}_{m}=F^{}_{m}$)
    and are given by Eq.~(\ref{Fm-ideal}). This is a natural generalization of the
     well-known von Neumann-L\"uders projection postulate for ideal quantum measurements
   (see, e.g., Ref.~\cite{BP-Book}). One can, however, construct more general measurement
   schemes associated with the notion of the positive operator-valued measure (POVM)~\cite{Holevo2001,SRM}. A POVM is defined by $N$ positive operators $F^{}_{m}$
    which form the resolution of the identity, but in general $F^{2}_{m}\not=F^{}_{m}$.
    Usually the operators $F^{}_{m}$ can be represented in the form (\ref{Fm-ideal}) where
    the set $\{|\psi^{}_{m}\rangle\}$ is \textit{overcomplete\/}, i.e., the number of
    outcomes $N$ exceeds the rank of the density matrix of the system
    (for a qubit $N>2$). In the present paper we will not consider such general situations
     and restrict ourselves to the von Neumann-L\"uders projection measurements.

\section{The dephasing model: dynamics of decoherence\label{secIII}}

 \subsection{Exact expressions for the coherences}
 Our central goal in this section is to demonstrate that
  the qubit dynamics with initial states prepared by non-selective
   measurements  exhibits a number of  physically interesting and even somewhat
    unexpected features when compared with the case of selective measurements.
   This is especially important when one is dealing with decoherence phenomena.

 Our discussion will be based on the analysis of the simple \textit{dephasing model\/}
 describing the main decoherence mechanism
 for certain types of system-environment  interactions~\cite{MMR2012,Luczka1,Unruh,MR-CMP2012,myCMP}.
 In this model, a two-state system (qubit) ($S$) is coupled to a bath ($B$) of
harmonic oscillators. Using the ``spin'' representation for the qubit with the basis states
 (\ref{qbit-can-bas}), the total Hamiltonian in the Schr\"odinger picture is taken to be
 (in our units $\hbar = 1$)
    \begin{eqnarray}
    \label{H}
    \nonumber
  &&  \hspace*{-20pt} H=H_S+H_B+H_{int}\\
   && \hspace*{-20pt} {}=\frac{\omega_0}{2}\sigma_3
   +\sum\limits_k\omega_k b^{\dagger}_k b_k
   +\sigma_3\sum\limits_k(g_k b^{\dagger}_k+g^*_k b_k),
   \end{eqnarray}
where $\omega_0$ is the energy difference between the excited
 state $|1\rangle$ and the ground state $|0\rangle$  of the qubit.
Bosonic creation and annihilation operators $b^{\dagger}_k$ and
$b_k$ correspond to the $k$th bath mode with frequency $\omega_k$,
and $g_k$ are the coupling constants.

Suppose that at time $t=0$ the state of the composite system ($S$+$B$) is characterized
by some density matrix $\varrho^{}_{SB}(0)$. Then at time $t$ the average value
 of a Heisenberg picture operator $A(t)$ is given by
 \begin{equation}
    \label{A(t)}
   \langle A(t)\rangle= \text{Tr}\left\{
     \exp(iHt)A\exp(-iHt)\varrho^{}_{SB}(0)
   \right\}.
  \end{equation}
 Below, the notation $\langle A\rangle$ will be used  for averages at $t = 0$.

 The quantities of principal interest are the
  \textit{coherences\/} $\langle \sigma^{}_{\pm}(t)\rangle$, where
   $\sigma^{}_{\pm}=\left(\sigma^{}_{1}\pm i\sigma^{}_{2}\right)/2$.
  They are related directly to
   the off-diagonal elements of the reduced density matrix of the qubit:
    \begin{equation}
       \label{coher-DM}
      \langle \sigma^{}_{+}(t)\rangle=\langle 0| \varrho^{}_{S}(t)|1\rangle,
      \qquad
       \langle \sigma^{}_{-}(t)\rangle= \langle 1| \varrho^{}_{S}(t)|0\rangle,
    \end{equation}
 where
   \begin{equation}
      \label{rho-S}
     \varrho^{}_{S}(t)= \text{Tr}^{}_{B}
     \left\{
          \exp(-iHt) \varrho^{}_{SB}(0)\exp(iHt)
     \right\}.
   \end{equation}

  The dephasing model (\ref{H}) has two distinctive features. First,
  the operator $\sigma^{}_{3}$ commutes with the Hamiltonian and,
   consequently, the average populations of the canonical  states (\ref{qbit-can-bas})
    do not depend on time. Thus we have a unique situation where the system relaxation may be interpreted physically as ``pure'' decoherence and exchange of entropy~\cite{MR-CMP2012,myCMP} rather than dissipation of energy. Second, in this model the equations of motion for
  all relevant operators can be solved exactly~\cite{MMR2012}. This allows one to
   study  the time evolution of the coherences for different initial
   conditions. Here we leave out many details for which we refer to
    Ref.~\cite{MMR2012} and simply quote some important results.

    If the initial state is prepared by a non-selective measurement, then,
    taking the initial density matrix of the composite system in the form
       (\ref{rho-SB-0}), we get for the coherences (\ref{coher-DM})
       \begin{equation}
    \label{coh1}
   \langle\sigma_{\pm}(t)\rangle=\frac{1}{Z}\sum_{m}\text{Tr}
   \left[\Omega_m^{\dagger}\sigma_{\pm}(t)\Omega_m e^{-\beta H}\right],
  \end{equation}
  where $\beta=1/k^{}_{\text{B}}T $, and $Z=\text{Tr}\left\{\exp(-\beta H)\right\}$
  is the equilibrium partition function. The analogous formula for the case of a selective
   measurement [see Eq.~(\ref{rho-SB-selec})] is evident.

    As shown in Ref.~\cite{MMR2012}, the time-dependent
    qubit operators $\sigma^{}_{\pm}(t)$ in the dephasing model (\ref{H}) are given by
       \begin{equation}
       \label{sig-pm-t}
       \sigma_{\pm}(t)= \exp\left\{ \pm i\omega_0 t \mp R(t) \right\} \sigma_{\pm}
       \end{equation}
   with
      \begin{equation}
    \label{R_alpha}
    R(t)=\sum_{k}\!\left[\alpha_{k}(t) b^{\dagger}_k-\alpha^*_k(t) b_{k} \right],
  \quad
  \alpha_k(t)\!=2g_{k}\frac{1-e^{i\omega_kt}}{\omega_{k}}.
 \end{equation}
Using the above expressions and the exact relations
   \begin{equation}
   \label{rho_relations}
   \begin{array}{l}
    e^{-\beta H}|0\rangle= e^{\beta\omega_0/2}e^{-\beta H^{(-)}_B}\otimes|0\rangle,\\[5pt]
     e^{-\beta H}|1\rangle= e^{-\beta\omega_0/2}e^{-\beta H^{(+)}_B}\otimes|1\rangle,
   \end{array}
   \end{equation}
where
  \begin{equation}
  \label{HBpm}
  H_B^{(\pm)}=\sum\limits_k\omega_k b^{\dagger}_k  b_k
  \pm\sum\limits_k(g_k b^{\dagger}_k+g^*_k b_k),
 \end{equation}
it is  a straightforward matter
to carry out the trace over the bath degrees of freedom in Eq.~(\ref{coh1}).
After some algebra  (for details see Ref.~\cite{MMR2012}), one obtains
  \begin{widetext}
      \begin{eqnarray}
      \label{coh22}
   && \hspace*{-2mm}
   \langle\sigma_{\pm}(t)\rangle=\langle\sigma_{\pm}\rangle\, e^{\pm i\omega_0 t} e^{-\gamma(t)}
       \frac{\sum_m\left\{\langle 0|\Omega_m^{\dagger}\sigma_{\pm}\Omega_m|0\rangle
   e^{\beta\omega_0/2\pm i\Phi(t)}
  +\langle 1|\Omega_m^{\dagger}\sigma_{\pm}\Omega_m|1\rangle
    e^{-\beta\omega_0/2\mp i\Phi(t)}\right\}}
    {\sum_m\left\{\langle 0|\Omega_m^{\dagger}\sigma_{\pm}\Omega_m|0\rangle
   e^{\beta\omega_0/2} +\langle 1|\Omega_m^{\dagger}\sigma_{\pm}\Omega_m|1\rangle
    e^{-\beta\omega_0/2}\right\}}
 \end{eqnarray}
  \end{widetext}
 with the initial coherences
   \begin{eqnarray}
     \label{sig-NonS-init}
     &&
     \hspace*{-20pt}
     \langle \sigma^{}_{\pm}\rangle =
     \frac{1}{2\cosh\left(\beta\omega^{}_{0}/2\right)}
     \sum_m \left\{
      \langle 0| \Omega^{\dagger}_{m}\sigma^{}_{\pm}\Omega^{}_{m}|0\rangle e^{\beta\omega_0/2}\right.
      \nonumber\\[2pt]
      && \hspace*{60pt}\left.
      {}+
      \langle 1| \Omega^{\dagger}_{m}\sigma^{}_{\pm}\Omega^{}_{m}|1\rangle e^{-\beta\omega_0/2}
      \right\}.
   \end{eqnarray}
 In the case of a selective measurement when the qubit is prepared in a pure quantum
  state $|\psi\rangle=c^{}_{0}|0\rangle + c^{}_{1} |1\rangle$
   with $|c^{}_{0}|^2 + |c^{}_{1}|^2=1$, the initial density matrix of the composite
    system is taken in the form (\ref{rho-SB-selec}).
    Then, instead of Eq.~(\ref{coh22}), we have
         \begin{eqnarray}
         \label{coh-selec}
     &&  \hspace*{-20pt}
     \langle\sigma_{\pm}(t)\rangle=\langle\sigma_{\pm}\rangle\,
      e^{\pm i\omega_0 t} e^{-\gamma(t)}\nonumber\\[2pt]
     && {}\times\frac{|c^{}_{0}|^2 e^{\beta\omega_0/2\pm i\Phi(t)}+
   |c^{}_{1}|^2 e^{-\beta\omega_0/2\mp i\Phi(t)}}{|c^{}_{0}|^2 e^{\beta\omega_0/2} +
   |c^{}_{1}|^2 e^{-\beta\omega_0/2}},
        \end{eqnarray}
where $\langle \sigma^{}_{\pm}\rangle=\langle \psi|\sigma^{}_{\pm}|\psi\rangle $.

 Formulas (\ref{coh22}) and (\ref{coh-selec}) contain two relevant functions. The so-called
  \textit{decoherence function\/} $\gamma(t)$ is defined as
  \begin{equation}
   \label{gamma-def}
   \gamma(t)=\int_0^{\infty}d\omega\, J(\omega)\coth(\beta\omega/2)
   \frac{1-\cos\omega t}{\omega^2},
   \end{equation}
  where the continuum limit of the bath modes is performed,
and the spectral density $J(\omega)$ is introduced by the rule
  \begin{equation}
     \label{J(omega)}
   \sum_{k} 4|g^{}_{k}|^2\,f(\omega^{}_{k})= \int_0^{\infty} d\omega\, J(\omega) f(\omega).
   \end{equation}
 It is clear that $\gamma(t)$ is precisely  the quantity which determines
  the relaxation of the off-diagonals (\ref{coher-DM})
  due to vacuum and thermal fluctuations in the bath~\cite{BP-Book}.
  The other function, $\Phi(t)$, is given by
    \begin{equation}
   \label{Phi}
  \Phi(t)=\int_0^{\infty}d\omega J(\omega) \frac{\sin\omega t}{\omega^2}.
   \end{equation}
As discussed in Ref.~\cite{MMR2012}, this function accounts for the influence of initial
 qubit-environment correlations on the dynamics of decoherence.
These correlations are inherited from the pre-measurement equilibrium
state due to the presence of the interaction term in the total
Hamiltonian (\ref{H}). Mathematically, it has the effect that
the operators $H^{+}_{B}$ and $H^{-}_{B}$ [see Eqs.~(\ref{rho_relations}) and
 (\ref{HBpm})] differ in the sign of the interaction term.

\subsection{Selective measurements}

  The result (\ref{coh-selec}) for selective measurements has been analyzed in detail in Ref.~\cite{MMR2012}, so that here we merely touch on some relevant points.

 The expression (\ref{coh-selec}) can be rewritten
 more transparently as
    \begin{equation}
     \label{coh3}
 \langle\sigma_{\pm}(t)\rangle=\langle\sigma_{\pm}\rangle
\exp[\pm i(\omega_0 t+\chi(t)] \exp[-\widetilde\gamma(t)],
    \end{equation}
where
  \begin{equation}
     \label{gamma-eff}
  \widetilde\gamma(t)=\gamma(t)+ \gamma^{}_{cor}(t)
    \end{equation}
is the effective decoherence function including the correlation
contribution
    \begin{equation}
    \label{gamma_cor}
     \gamma_{cor}(t)\!= -\frac{1}{2}\ln\!
     \left[
        1\!-\frac{(1-\langle\sigma_3\rangle^2)\sin^2\Phi(t)}
            {[\cosh(\beta\omega_0/2)-\langle\sigma_3\rangle\sinh(\beta\omega_0/2)]^2}
    \right],
 \end{equation}
whereas $\chi(t)$ is the time-dependent phase shift with
  \begin{equation}
  \label{chi}
  \tan\chi(t)=\frac{\sinh(\beta\omega_0/2)-
 \langle\sigma_3\rangle\cosh(\beta\omega_0/2)}
 {\cosh(\beta\omega_0/2)-\langle\sigma_3\rangle\sinh(\beta\omega_0/2)}\tan\Phi(t).
 \end{equation}
In writing the above formulas we have used the obvious relations
 $\langle \sigma^{}_{3}\rangle=\langle \psi|\sigma^{}_{3}|\psi\rangle=
  |c^{}_{1}|^2 - |c^{}_{0}|^2$.

It is important to note and  easy to see from Eqs.~(\ref{gamma-def}) and (\ref{gamma_cor})
that both terms in the effective decoherence function (\ref{gamma-eff})
are always \textit{positive\/}. Thus, in cases where the initial state
 is prepared by a selective measurement, initial qubit-bath correlations may be
 viewed as an additional  source of decoherence.

\subsection{Non-selective measurements}

 Now we turn to Eq.~(\ref{coh22})  and consider the general
 non-selective measurement scheme described by Eqs.~(\ref{QB-b12-FOmega}).
  Using the representation (\ref{a-qubit}), we obtain after some algebra
   the expression (\ref{coh3}) in which the correlation part of the decoherence function is
    now given by
 \begin{eqnarray}
 \label{gam_cor_NS}\nonumber
   \hspace*{-15pt}\gamma_{cor}(t)=-\frac{1}{2}\ln\left\{
  1+\left(\frac{N_1^2+N_2^2}{D^2}-1\right)\sin^2\Phi(t)\right.
  \\
   \hspace*{-15pt}\left.{}+\frac{N_2}{D}\sin(2\Phi(t))\right\},
 \end{eqnarray}
 where we have introduced
\begin{widetext}
 \begin{eqnarray}
 \label{NND}
 \nonumber
  && \hspace*{-11pt}
  N_1\!=\!\left\{e^{\beta\omega_0}\sin^4(\theta_a/2)-e^{-\beta\omega_0}\cos^4(\theta_a/2)
  \right\}\sin^2\theta_1+
    \left\{e^{\beta\omega_0}\cos^4(\theta_a/2)-e^{-\beta\omega_0}\sin^4(\theta_a/2)
  \right\}\sin^2\theta_2
 \\
\nonumber
 && \hspace*{-11pt}
 \quad\quad{} +\sinh(\beta\omega_0)\sin^2\theta_a\cos\Delta_{\phi}\sin\theta_1\sin\theta_2,
  \\[1ex]
 &&   \hspace*{-11pt}
   N_2\!=\!2\cos\theta_a\,\sin\Delta_{\phi}\, \,\sin \theta_1\, \sin \theta_2,
  \\[1ex]
  \nonumber
  && \hspace*{-11pt}
  D\!=\!\left\{\mbox{$\frac{1}{2} $}
  \sin^2\theta_a +e^{\beta\omega_0}\sin^4(\theta_a/2)+e^{-\beta\omega_0}\cos^4(\theta_a/2)
  \right\}\sin^2\theta_1 \!+\!
 \left\{\mbox{$\frac{1}{2} $}
 \sin^2\theta_a +e^{\beta\omega_0}\cos^4(\theta_a/2)+e^{-\beta\omega_0}\sin^4(\theta_a/2)
  \right\}\sin^2\theta_2
\\
  \nonumber
  && \hspace*{-10pt}
  \quad {}+\left\{\cosh(\beta\omega_0)\sin^2\theta_a
   +2\left[\sin^4(\theta_a/2)+\cos^4(\theta_a/2)\right]
       \right\}\cos\Delta_{\phi}\sin\theta_1\sin\theta_2
  \end{eqnarray}
  \end{widetext}
 To simplify notation, we have written $\theta^{}_{i}$,  $\phi^{}_{i}$ for
 $\theta^{}_{b_i}$, $\phi^{}_{b_i}$, and denoted $\Delta_{\phi}=\phi_1-\phi_2$.
 The expression for the phase shift $\chi(t)$ is
 \begin{equation}
 \label{chi_NS}
 \chi(t)=\arctan\left(\frac{N_1\sin\Phi(t)}{D\cos\Phi(t)+N_2\sin\Phi(t)}\right).
 \end{equation}
The initial coherences (\ref{sig-NonS-init}) can be directly evaluated to yield
  \begin{eqnarray}
    \label{sig0-2}
    & & \hspace*{-28pt}
   \langle\sigma_{\pm}\rangle\!=\!\frac{e^{\pm i\phi^{}_{1}}}{4\cosh(\beta\omega_0/2)}
  \nonumber\\[2pt]
  & & {}\times
   \left\{ \sin\theta^{}_{1}\!\left[ e^{\beta\omega_0/2}
     \sin^2\frac{\theta_a}{2}  +e^{-\beta\omega_0/2}\cos^2\frac{\theta_a}{2}\right] \right.
   \nonumber\\[2pt]
     & & \hspace*{-20pt}
     \left. {}\!+\!{e}^{\mp i\Delta_{\phi}}\sin\theta^{}_{2}
      \!\left[ e^{\beta\omega_0/2} \cos^2\frac{\theta_a}{2}
     \!+\! e^{-\beta\omega_0/2}  \sin^2\frac{\theta_a}{2}\right] \right\}.
  \end{eqnarray}
Formulas (\ref{gam_cor_NS})\,--\,(\ref{chi_NS}), together with Eqs.~(\ref{coh3})
and (\ref{gamma-eff}), determine  the time-dependent coherences
 $\langle\sigma^{}_{\pm}(t)\rangle $ or, what is the same, the off-diagonals of the
  qubit density matrix [see Eq.~(\ref{coher-DM})].
   It is often convenient to use the representation of the qubit density matrix
 $\varrho^{}_{S}(t)$ in terms of the Bloch vector
  $\vec{v}(t)=\langle\vec{\sigma}(t)\rangle$~\cite{22inPRA12,23inPRA12}:
    \begin{equation}
           \label{rho-v}
           \begin{array}{l}
    \varrho^{}_{S}(t)=\frac{1}{2}\left[ 1+ \vec{\sigma}\cdot\vec{v}(t)\right].
           \end{array}
   \end{equation}
The magnitude  of the Bloch vector satisfies $0\leq v(t)\leq 1$, and
        $v=1$ only if the qubit is in a pure quantum state. It is easy to check
         that
   \begin{equation}
          \label{Bloch}
   v(t)=\left[4\langle\sigma_+(t)\rangle \langle\sigma_-(t)\rangle
   +\langle\sigma_3(t)\rangle^2\right]^{1/2}.
   \end{equation}
   Since $\sigma^{}_{3}$ commutes with the Hamiltonian (\ref{H}), we have
     $\langle\sigma_3(t)\rangle=\langle\sigma_3\rangle$.
     Taking also into account Eq.~(\ref{coh3}), we obtain
        \begin{equation}
      \label{Bloch-expl}
       v(t)=\left[4\langle\sigma_{+}\rangle
         \langle\sigma_{-}\rangle {e}^{-2\widetilde{\gamma}(t)}
         +\langle\sigma_3\rangle^2 \right]^{{1}/{2}}.
        \end{equation}
Thus, to calculate $v(t)$, all we need is the initial average $\langle\sigma^{}_{3}\rangle$.
For the general non-selective measurement scheme (\ref{QB-b12-FOmega}), this
 average can be found by using the formula analogous to (\ref{sig-NonS-init}).
 A straightforward algebra gives
   \begin{eqnarray}
   \label{sigZ}
   \nonumber & & \hspace*{-30pt}
   \langle\sigma_{3}\rangle= \frac{1}{2\cosh(\beta\omega_0/2)}
   \\[2pt]
   \nonumber & & \hspace*{-20pt}
    {}\times\left\{\cos\theta_1\left[ e^{\beta\omega_0/2}\sin^2\frac{\theta_a}{2}
   + e^{-\beta\omega_0/2}\cos^2\frac{\theta_a}{2}\right]\right.
   \\[2pt]
   & & \hspace*{-10pt} \left.
   {}+\cos\theta_2 \left[e^{\beta\omega_0/2}\cos^2\frac{\theta_a}{2}
    + e^{-\beta\omega_0/2}\sin^2\frac{\theta_a}{2}\right]\right\}.
 \end{eqnarray}

\subsection{Some special non-selective preparation measurements}

To gain an insight into new features of the qubit dynamics in cases when
the initial state is prepared by a non-selective measurement, we will apply
 the above general expressions to some special preparation schemes described
 in Sec.~\ref{secII}.

 We start with the scheme~(\ref{QB-b-FOmega}) for which,
 according to relations~(\ref{Euler;-a}),
    \begin{equation}
   \label{theta-phi-1}
   \theta^{}_{1}+\theta^{}_{2}=\pi,
   \quad
   \sin \Delta^{}_{\phi}=0,
   \quad
   \cos\Delta^{}_{\phi}=-1,
  \end{equation}
 where $\theta^{}_{1}\equiv \theta^{}_{b}$ and $\phi^{}_{1}\equiv \phi^{}_{b}$.
 Noting that in this case $N^{}_{2}=0$, Eq.~(\ref{gam_cor_NS})
is manipulated to the simple  form
    \begin{equation}
   \label{gam_cor_NS1}
    \gamma_{cor}(t)=
    -\frac{1}{2}\ln\left[1+\frac{\sin^2\Phi(t)}{\sinh^2(\beta\omega_0/2)} \right].
    \end{equation}
    For the phase shift (\ref{chi_NS}) we find
    \begin{equation}
    \label{shift}
    \tan\chi(t)=\coth(\beta\omega_0/2)\tan\Phi(t).
    \end{equation}
   The results (\ref{gam_cor_NS1}) and (\ref{shift}) have several notable  properties. First,  they are  \textit{universal\/} in the  sense that they  do not depend on the qubit states
     $|\vec{a}\rangle$ and $|\vec{b}\rangle$  in Eqs.~(\ref{QB-b-FOmega}) describing  this
      type of non-selective measurements. In particular, the same expressions for
       $ \gamma_{cor}(t)$ and $\chi(t)$ hold for the simplest scheme
       (\ref{Uunity}) where $|\vec{b}\rangle=|\vec{a}\rangle$, i.e., the measuring device does not disturb the basis states. Next, we note that the function
        (\ref{gam_cor_NS1}) satisfies $\gamma_{cor}(t)\leq 0$
         at all times $t$. In other words, we have an \textit{enhancement\/}
        of coherence in the qubit caused by initial qubit-bath correlations!
         Moreover, it is seen from Eq.~(\ref{gam_cor_NS1}), that the $|\gamma_{cor}(t)|$ grows with temperature, so that in the temperature range $\beta\omega^{}_{0}\ll 1$ the effective decoherence function (\ref{gamma-eff}) may even become negative, at least
         during the initial stage of the system's evolution. To illustrate this point, we have evaluated the reduced coherence
           \begin{equation}
             \label{red-coh}
             |\langle\sigma(t)\rangle|/|\langle\sigma\rangle|\equiv
             |\langle\sigma_{\pm}(t)\rangle|/|\langle\sigma_{\pm}\rangle|=
             \exp\left[-\widetilde{\gamma}(t)\right]
           \end{equation}
 using Eqs.~(\ref{gamma-def}), (\ref{gamma-eff}), and (\ref{gam_cor_NS1}).
  The bath spectral density was taken in the form
         \begin{equation}
          \label{spectral}
          J(\omega)=\lambda_s\,\omega_c^{1-s}\omega^s e^{-\omega/\omega_c},
         \end{equation}
 which is most commonly used in the theory of spin-boson
systems \cite{BP-Book,Luczka1,Unruh,Leggett}.
 Here $\omega_c$ stands for some ``cutoff'' frequency,
and $\lambda_s$ is a dimensionless coupling constant.
The ``dynamical part'' (\ref{gamma-def}) of the decoherence function and the function
 $\Phi(t)$, Eq.~(\ref{Phi}), have been studied in detail in Ref.~\cite{MMR2012} for the
  sub-Ohmic ($0<s<1$), Ohmic ($s=1$), and super-Ohmic ($s>1$) cases. Here we shall restrict
  ourselves to the most prominent Ohmic case where
    \begin{equation}
      \label{PhiOm}
      \Phi(t)=\lambda\,\arctan(\omega_c t),\qquad\lambda\equiv\lambda_1,
    \end{equation}
   and the time behavior  of the correlation   term (\ref{gam_cor_NS1}) is very sensitive to the value of the
   coupling constant $\lambda$.
 \vspace{0.5cm}
 \begin{figure}[htb]
\centerline{\includegraphics[height=0.3\textheight,angle=0]{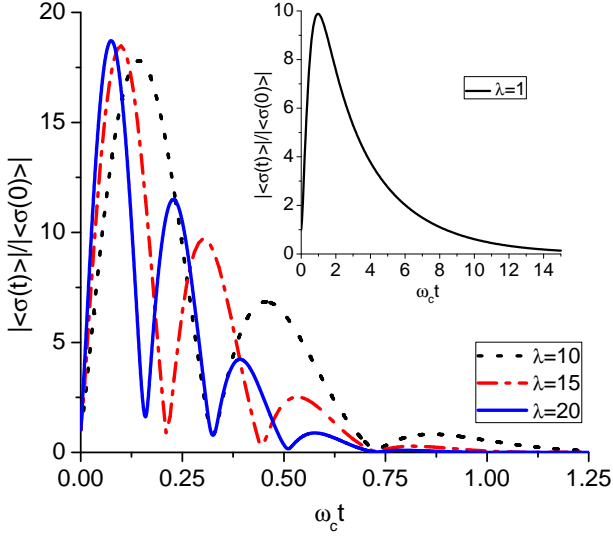}}
\vspace{0.2cm}
 \caption{Time evolution of the reduced coherence
in the Ohmic case for different values of the dimensionless coupling
constant. The preparation measurement is described by Eqs.~(\ref{QB-b-FOmega}).
Parameter values:
$\beta\omega_0=0.1$, $\omega_0/\omega_c=0.01$.}
\label{Fig-Sigma}
  \end{figure}

 Figure~\ref{Fig-Sigma} shows the time dependence
    of the reduced coherence in the temperature
    range $\beta\omega^{}_{0}\ll 1$. At the initial stage of evolution
    ($ \omega^{}_{c}t\lesssim 1$) the correlation effects dominate, so that
     the decoherence function (\ref{gamma-eff}) takes negative values.  At times
      $ \omega^{}_{c}t \gg 1$ the correlation effects are suppressed due
       to vacuum and thermal fluctuations contributing to the
       ``dynamical part'' (\ref{gamma-def}) of the decoherence function.
       Using the explicit expressions for the function (\ref{Phi})
       in the super-Ohmic ($s>1$) and sub-Ohmic ($0<s<1$) cases
       (see Ref.~\cite{MMR2012}), it can
       be shown that a similar time behavior of the reduced coherence is expected
        in these coupling regimes.

             It is worthwhile remarking that even for a moderate
    qubit-bath coupling ($\lambda\approx 1$), the maximum value of the coherences
    $|\langle \sigma^{}_{\pm}(t)\rangle|$ may be by one order of magnitude larger than the initial value  $|\langle \sigma^{}_{\pm}\rangle|$. This fact appears at first rather strange
    and even paradoxical. In particular, it seems likely that for some
      initial conditions, the magnitude of the Bloch  vector~(\ref{Bloch-expl}) may
     exceed the unity. This is, of course, not so for the following reason.
     The point is that the initial averages (\ref{sig0-2}) and (\ref{sigZ}) depend on temperature. For the preparation procedure under consideration
     [see Eqs.~(\ref{theta-phi-1})], we find
      \begin{equation}
        \label{sig-init-I}
         \begin{array}{l}
           \langle\sigma^{}_{\pm}\rangle=- \frac{1}{2} {e}^{\pm i\phi_b}
           \tanh\left(\beta\omega_0/2\right)\cos\theta^{}_a\sin\theta^{}_{b} ,
           \\[5pt]
           \langle \sigma_3\rangle=
            -\tanh\left(\beta\omega_0/2\right)\cos\theta^{}_a\cos\theta^{}_{b} .
         \end{array}
      \end{equation}
  In the temperature range $\beta\omega^{}_{0}\ll 1$, we have
  $|\langle\sigma^{}_{\pm}\rangle|\ll 1$ and
  $|\langle\sigma^{}_{3}\rangle|\ll 1$. Recalling Eqs.~(\ref{Bloch-expl})
  and (\ref{gam_cor_NS1}), one can show  that, at all times, $v(t)\leq 1$,
  as it must be. Other physical questions related to the enhancement of coherence
  for this preparation scheme  will be discussed in Sec.~\ref{secIV}.

    Now we shall consider the dynamics of decoherence in the case of another
     notable  non-selective preparation scheme described by Eqs.~(\ref{Omega-non}).
     This scheme corresponds to the following choice in Eqs.~(\ref{NND}):
          \begin{equation}
   \label{theta-phi-2}
   \theta^{}_{1}=\theta^{}_{2}\equiv \theta^{}_{b},
   \quad
   \sin \Delta^{}_{\phi}=0,
   \quad
   \cos\Delta^{}_{\phi}=1.
  \end{equation}
  From the second of Eqs.~(\ref{NND}) it is clear that $N^{}_{2}$ is again zero.
 Then, after some simple algebra Eqs.~(\ref{gam_cor_NS}) and  (\ref{chi_NS}) give
    \begin{eqnarray}
    \label{gam_cor_NS-NON}
      & & \gamma_{cor}(t)=
     -\frac{1}{2}\ln\left[1-\frac{\sin^2\Phi(t)}{\cosh^2(\beta\omega_0/2)}\right],
     \\[5pt]
     \label{shift-NON}
      & & \tan\chi(t)=\tanh(\beta\omega_0/2)\tan\Phi(t).
      \end{eqnarray}
 The initial averages (\ref{sig0-2}) and (\ref{sigZ}) now take the form
    \begin{equation}
       \label{sig-init-II}
          \begin{array}{l}
           \langle\sigma^{}_{\pm}\rangle=\frac{1}{2} {e}^{\pm i\phi_b}\sin\theta^{}_{b},
           \qquad
           \langle \sigma_3\rangle=\cos\theta^{}_{b}.
           \end{array}
      \end{equation}

   Similar to formulas~(\ref{gam_cor_NS1}) and (\ref{shift}),
  the results (\ref{gam_cor_NS-NON}) and (\ref{shift-NON}) are \textit{universal\/}
  in the sense that they do not depend on the qubit states
   $|\vec{a}\rangle$ and  $|\vec{b}\rangle$ in Eqs.~(\ref{Omega-non}).
    Another distinctive property  of Eq.~(\ref{gam_cor_NS-NON}) is that
    $\gamma^{}_{cor}(t)\geq 0$ at all times. Physically, in this case
     the initial qubit-bath correlations lead to additional decoherence.
     It is also interesting to note that expressions (\ref{gam_cor_NS-NON}) and  (\ref{shift-NON})  are identical to expressions (\ref{gamma_cor}) and (\ref{chi}) for the \textit{selective\/} measurement with  $\langle \sigma^{}_{3}\rangle=0$, i.e., with equal populations of the basis states (\ref{qbit-can-bas}). This is not accidental; for a discussion see Appendix~\ref{App-pure-nonsel}.

  A comparison of Eqs.~(\ref{theta-phi-1}) and (\ref{theta-phi-2}), together with the fact that the Euler angles $\theta^{}_{1}$ and $\theta^{}_{2}$ enter the functions (\ref{NND}) only through $\sin\theta^{}_{1}$ and $\sin\theta^{}_{2}$, suggests that
 $\Delta^{}_{\phi}=\phi^{}_{1}-\phi^{}_{2}$ is a key quantity determining the
  main qualitative features of the system's evolution.
 To illustrate this point, suppose that the system is initially
   prepared by the non-selective measurement
    (\ref{QB-b12-FOmega}) with the Euler angles of
    the qubit states $|\vec{b}^{}_{1}\rangle$ and $|\vec{b}^{}_{2}\rangle$ satisfying
            \begin{equation}
   \label{theta-phi-3}
   \theta^{}_{1}=\theta^{}_{2},
   \quad
   \sin \Delta^{}_{\phi}=0,
   \quad
     \cos\Delta^{}_{\phi}=-1.
           \end{equation}
The corresponding initial averages (\ref{sig0-2}) and (\ref{sigZ}) are
       \begin{equation}
        \label{sig-init-III}
         \begin{array}{l}
           \langle\sigma^{}_{\pm}\rangle=- \frac{1}{2} {e}^{\pm i\phi_1}
           \tanh\left(\beta\omega_0/2\right)\cos\theta^{}_a\sin\theta^{}_{1} ,
           \\[5pt]
           \langle \sigma_3\rangle=\cos\theta^{}_{1}.
         \end{array}
      \end{equation}
 We mention that formulas (\ref{theta-phi-3}) differ from Eqs.~(\ref{theta-phi-2})
  only in that $\cos\Delta^{}_{\phi}$ is now of opposite sign.  Nevertheless, it is evident
  that we obtain for $\gamma^{}_{cor}(t)$ the result (\ref{gam_cor_NS1})
  which corresponds to the  entirely different evolution of the coherences
   $\langle \sigma^{}_{\pm}(t)\rangle$.

     It would be instructive to look at the last
    example from another point of view. Let us write the post-measurement qubit state
     $|\vec{b}^{}_{1}\rangle$ in the canonical basis (\ref{qbit-can-bas}):
      \begin{equation}
         \label{b1-ampl}
      |\vec{b}^{}_{1}\rangle= c^{}_{0} |0\rangle + c^{}_{1} |1\rangle,
      \end{equation}
    where the amplitudes $ c^{}_{0}$ and $ c^{}_{1}$ can be expressed in terms
     of the Euler angles by using Eq.~(\ref{a-qubit}). Then
      using Eqs.~(\ref{theta-phi-3}) leads to the following representation for the state
      $|\vec{b}^{}_{2}\rangle $:
            \begin{equation}
         \label{b2-ampl}
      |\vec{b}^{}_{2}\rangle= i\big( c^{}_{0} |0\rangle - c^{}_{1} |1\rangle \big).
      \end{equation}
 There is no new physics in the appearance of $i$, but the additional phase shift
  between the basis states, as compared to Eq.~(\ref{b1-ampl}),
  radically influences the qubit's dynamics.

 Thus far we have been concerned  with special types of non-selective
  measurement schemes which lead to physically important features of decoherence.
    To give a comprehensive review of all possible regimes of evolution,
   one should appeal to Eq.~(\ref{gam_cor_NS}).  In general, the quantities (\ref{gam_cor_NS})--(\ref{chi_NS}) are rather complicated functions of the polar angles $\theta^{}_a$, $\theta^{}_1$, $\theta^{}_2$, and the
   difference $\Delta_{\phi}=\phi^{}_{1}-\phi^{}_{2}$ of the azimuthal angles.
   In addition, they depend on temperature and the qubit energy $\omega^{}_{0}$.
This makes a detailed analysis of $\gamma^{}_{cor}(t)$ rather
 cumbersome for our discussion. Nevertheless, we can formulate a simple sufficient
 condition that the initial preparation of the system by a general non-selective
 measurement (\ref{QB-b12-FOmega}) leads to enhancement of coherence in the qubit.
     Notice that for all qubit states $|\vec{b}^{}_{1}\rangle$ and
  $|\vec{b}^{}_{2}\rangle$ with $\sin\Delta^{}_{\phi}=0$, we have $N^{}_{2}=0$ and,
   consequently, the term with $\sin(2\Phi(t))$ in Eq.~(\ref{gam_cor_NS})  vanishes.
 Then we arrive at the conclusion that $\gamma^{}_{cor}(t)\leq 0$ at all times if
   \begin{equation}
        \label{Suff-gamma}
     N^{2}_{1} > D^2,
     \qquad
     \sin\Delta^{}_{\phi}=0.
   \end{equation}
 Clearly, the second condition implies $\cos\Delta^{}_{\phi}=\pm 1$.
 We will not give here a somewhat lengthy formal analysis of Eqs.~(\ref{Suff-gamma})
 since it does not add anything substantially new to the results of the above
 discussion. We only note that the enhancement of coherence takes place for
  the post-measurement states $|\vec{b}^{}_{1}\rangle$ and
   $|\vec{b}^{}_{2}\rangle$ with $\cos\Delta^{}_{\phi}=-1$ and
   $\theta^{}_{1}+\theta^{}_{2}\approx \pi$ or
   $\theta^{}_{1}\approx\theta^{}_{2}$. In other words, the post-measurement
    states should be close to the states in the measurement schemes
     (\ref{theta-phi-1}) or (\ref{theta-phi-3}).

  \section{The purity and entropy of the qubit \label{secIV}}

 If at time $t$  the effective decoherence function~(\ref{gamma-eff}) is negative, the state
  of the qubit is ``less mixed'' than initially. This property can be characterized
  quantitatively by  the von-Neumann-Shannon information entropy
    \begin{equation}
         \label{S-def}
     S(t)= - \text{Tr}^{}_{S}\left\{
        \varrho^{}_{S}(t)\,\ln\varrho^{}_{S}(t)
      \right\}.
    \end{equation}
Using Eq.~(\ref{rho-v}), the information entropy of a qubit can be expressed in terms of the Bloch vector magnitude~\cite{MMR2012}:
     \begin{equation}
     \label{S-v}
   S(t)= \ln 2
  - \frac{1}{2}\left(1+v\right)\ln\left(1+v\right)
  - \frac{1}{2}\left(1-v\right)\ln\left(1-v\right).
  \end{equation}
 Since $0\leq v \leq 1$, we have $0\leq S\leq \ln 2$ with $S=0$ for a pure quantum
  state ($v=1$).

  Another measure of the ``mixedness'' (or the lack of information about a system)
   is the so-called \textit{purity\/} of the system's state~\cite{Luczka,[2]inDajka}:
   \begin{equation}
    {\mathcal P}(t)= \text{Tr}^{}_{S}\left\{\varrho^{2}_{S}(t)\right\}.
   \end{equation}
  Again using Eq.~(\ref{rho-v}), we obtain for a qubit
    \begin{equation}
       \label{P-v}
    {\mathcal P}(t)= \frac{1}{2}\left( 1+v^2 \right).
    \end{equation}
  Obviously $1/2\leq {\mathcal P}\leq 1$ with ${\mathcal P}=1$ for a pure state.

  In principle, either  $S(t)$ or ${\mathcal P}(t)$
   may be used to measure the degree of coherence in a qubit.
   In both cases the key quantity is the magnitude of the Bloch vector (\ref{Bloch-expl}).
   Here we shall discuss  the time behavior of the purity and entropy in the
   non-selective  measurement schemes (\ref{theta-phi-1}) and (\ref{theta-phi-3}) for which
   the enhancement of coherence is most pronounced.

\begin{figure}[htb]
\centerline{\includegraphics[height=0.29\textheight,angle=0]{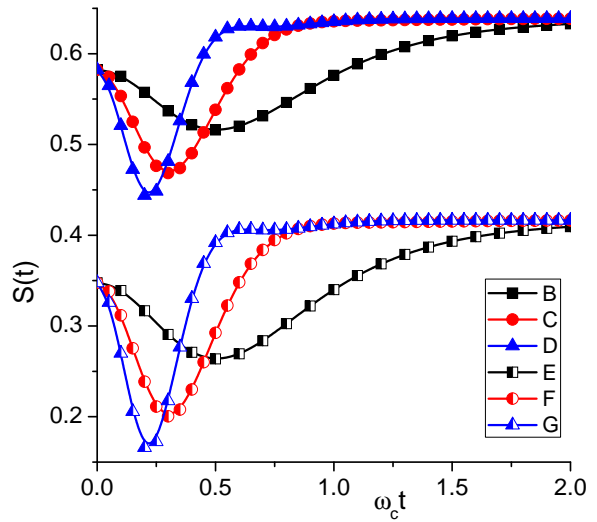}}
 \caption{Time evolution of the qubit entropy $S(t)$
in the Ohmic case for different values of the
coupling constant: $\lambda=2$ (B and E),  $\lambda=4$ (C and F),
 $\lambda=6$ (D and G). Filled symbols correspond to the preparation measurement described by
Eqs.~(\ref{theta-phi-1}), half-filled -- to the measurement
described by Eqs.~(\ref{theta-phi-3}). Other parameter values:
$\beta\omega_0=1$, $\omega_0/\omega_c=0.1$, $\theta_a=0$,
$\theta_1=\pi/4$.} \label{Slambda}
  \end{figure}

\begin{figure}[htb]
\centerline{\includegraphics[height=0.29\textheight,angle=0]{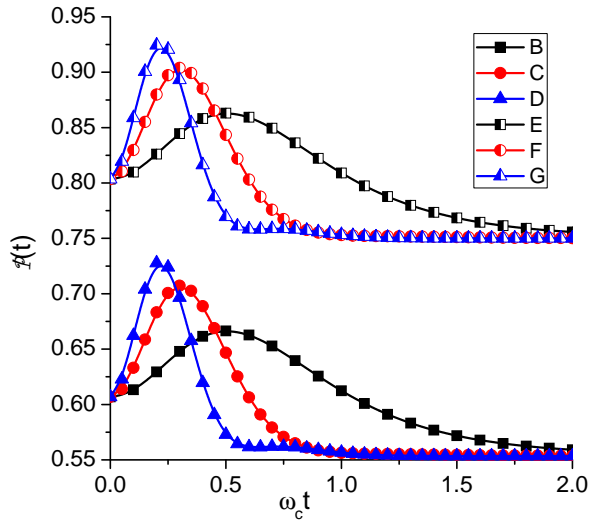}}
 \caption{Time evolution of the qubit purity ${\mathcal P}(t)$
in the Ohmic case for different values of the
 coupling constant. The symbols and the system parameters are
the same as in Fig.~\ref{Slambda}.} \label{Plambda}
  \end{figure}

 Figures~\ref{Slambda} and \ref{Plambda} illustrate the
evolution of the qubit entropy (\ref{S-v}) and purity (\ref{P-v})
at a fixed temperature for different values of the coupling
constant.

 It is seen that  with increasing the coupling constant the maximum purity
 becomes larger and is shifted to smaller $t$. For large values of the coupling constant,
  there appear oscillations in both the purity and entropy (cf. Fig.~\ref{Fig-Sigma}).
  According to Eqs.~(\ref{sig-init-I}) and (\ref{sig-init-III}),
   the initial coherences $|\langle\sigma^{}_{\pm}\rangle|$ are identical in the preparation
    measurement schemes  described by Eqs.~(\ref{theta-phi-1}) and (\ref{theta-phi-3}).
    Note also that the time dependence of the decoherence function
     $\widetilde{\gamma}(t)$ is the same in both cases.  The corresponding pairs of lines
     in Fig.~\ref{Plambda} are therefore just shifted vertically from each other
    because of the different values of $\langle\sigma_3\rangle$.
     Expressions (\ref{sig-init-I}) and (\ref{sig-init-III}) for $\langle\sigma_3\rangle$ show that, at all temperatures, the preparation scheme (\ref{theta-phi-1}) leads to less pure states of the qubit as compared to the scheme (\ref{theta-phi-3}).

\begin{figure}[htb]
\centerline{\includegraphics[height=0.3\textheight,angle=0]{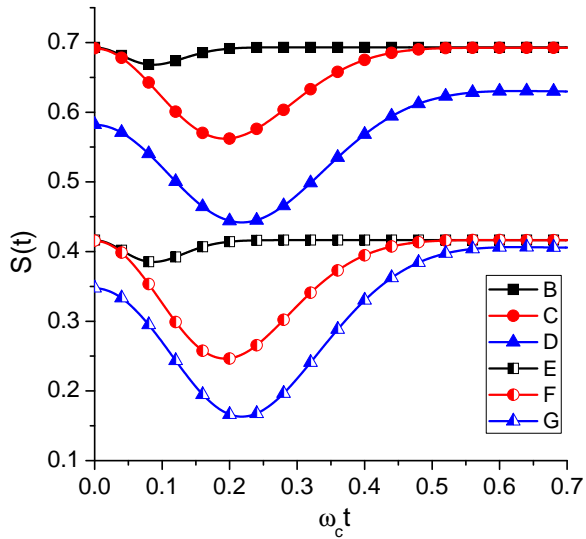}}
 \caption{Time evolution of the qubit entropy $S(t)$
in the Ohmic case at different temperatures:
$\beta\omega_0=0.01$ (B and E), $\beta\omega_0=0.1$ (C and F),
 $\beta\omega_0=1$ (D and G). Filled symbols correspond to the preparation
 measurement described by
Eqs.~(\ref{theta-phi-1}), half-filled -- to the measurement
described by Eqs.~(\ref{theta-phi-3}). Other parameter values:
$\lambda=6$, $\omega_0/\omega_c=0.1$, $\theta_a=0$,
$\theta_1=\pi/4$.} \label{Sbeta}
  \end{figure}

\begin{figure}[htb]
\centerline{\includegraphics[height=0.3\textheight,angle=0]{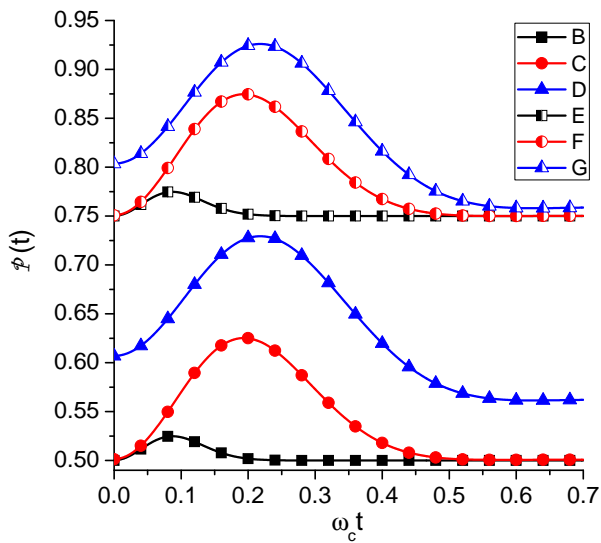}}
 \caption{Time evolution of the qubit purity ${\mathcal P}(t)$
in the Ohmic case at different temperatures. The symbols and the system parameters are
the same as in Fig.~\ref{Sbeta}.} \label{Pbeta}
  \end{figure}

Figures~\ref{Sbeta} and \ref{Pbeta}  display the evolution of the qubit purity and
entropy at different temperatures but for a fixed value of the coupling constant.

  It is seen that lowering of temperature
  is favourable for purification of the qubit state.
  To explain this, it should be remembered that the time evolution
   of the purity ${\mathcal P}(t)$ and the entropy $S(t)$ is the net result
   of two processes. First, there is the enhancement of coherence due to the
    combined effect of the non-selective preparation measurement and qubit-bath
    correlations. Second, thermal fluctuations in the bath
     suppress the coherence effects.
     At high temperatures, the latter process dominates. Note in this connection
     that  the maximum of the purity shifts to smaller $t$ with increasing
      temperature (see Fig.~\ref{Pbeta}).

\section{Conclusions\label{secV}}

In this paper we have examined the properties of the reduced qubit dynamics in cases where
 the initial state of the composite  system (qubit plus environment) is prepared
through non-selective quantum measurements. Our main result is that
 for some preparation schemes  the interplay of
 the measurement process and qubit-environment correlations
 can lead to a  significant enhancement of coherence in the qubit during the initial stage
  of evolution. The non-trivial feature of this effect is that, in general, a non-selective
   measurement  produces a \textit{mixed\/} state of the composite system.
   Then the purity of the qubit's state grows while its entropy  decreases with time
   until thermal fluctuations suppress this environmentally induced ``purification''
    process. It deserves to be pointed out that the temperature dependence of the
     purity growth is determined by several factors.
     First, the initial coherences
     (i.e., the off-diagonal elements of the qubit density matrix)  generated
       by a non-selective measurement decrease dramatically with increasing
     temperature. Second, the destructive effect of thermal fluctuations in the bath
      also becomes important just at high temperatures.
     But surprisingly, the dynamical enhancement of coherence due
      to the qubit-bath correlations grows with temperature, so that
       the maximum of the coherences $|\langle \sigma^{}_{\pm}(t)\rangle|$ may be much
        larger than the initial values $|\langle \sigma^{}_{\pm}\rangle|$
       (see Sec.~\ref{secIII}). Due to the interplay of the above factors,
        the resulting purity of the qubit states decreases with temperature, but not so
        rapidly as one might expect from intuitive considerations.

 Summarizing, if the initial state is prepared by a non-selective  measurement,
 it is possible to achieve the environmentally induced purity growth in a qubit  during some time interval by setting the measurement device in a proper way. This effect may be of interest
 in view of its connection with problems of measurement-based quantum control
 in open quantum systems. We refer, e.g., to the recent paper~\cite{Chaudhry14} were
the dephasing model (\ref{H}) was used to study the influence of system-environment correlations
 on the so-called quantum Zeno and anti-Zeno effects in repeated
 \textit{selective\/} measurements on single-qubit and many-qubit systems.

As a final remark we wish to emphasize that our study of the
environmentally induced purity growth does not claim to be
complete even on the single-qubit level.
 Here we will touch briefly on two open questions which
  deserve further investigation. First, our analysis was based
   on the assumption that initial correlations between the qubit and the environment
 are inherited from the premeasurement equilibrium
state due to the presence of the interaction term in the total
Hamiltonian $H$.  This assumption is adequate in describing many
 real situations and is commonly accepted in the theory of open quantum
  systems~\cite{2inPRA12,14inPRA12,18inPRA12}. Nevertheless, one may
   imagine a variety of different correlated initial states
    by replacing  $\varrho^{}_{\rm eq}$ in Eq.~(\ref{rho-SB-0})
    by some  nonequilibrium premeasurement density matrix $\varrho^{\prime}$.
     One should note, however, that within such a general formulation of the
      problem the trace over the bath degrees of freedom in
      $\langle \sigma^{}_{\pm}(t)\rangle $ cannot be carried out without
       a detailed information about
     physically reasonable forms of $\varrho^{\prime}$.
     The second important point is that  the model (\ref{H}) describes only the dephasing mechanism
of decoherence. Although this mechanism can dominate in real
physical systems~\cite{24inPRA12,25inPRA12}, in general the
population decay should be
 taken into account. However, it is a challenging problem because
  in this case the model is no longer exactly solvable.

\appendix

\section{Preparation of pure qubit states through non-selective measurements \label{App-pure-nonsel}}

\renewcommand{\theequation}{B.\arabic{equation}}
\setcounter{equation}{0}

Here we briefly discuss the connection between the non-selective preparation
 scheme (\ref{Omega-non}) [see also Eqs.~(\ref{theta-phi-2})] and selective measurements.

Using the explicit expressions for the $\Omega$-operators, the initial density
 matrix (\ref{rho-SB-0}) of the composite system  is written as
    \begin{equation}
       \label{A:Init-non}
      \varrho^{}_{SB}(0)=|\vec{b}\rangle \langle \vec{b}|\otimes \varrho^{}_{B}(\vec{a})
     \end{equation}
with the bath density matrix
        \begin{equation}
    \label{A:B-init-1}
    \varrho^{}_{B}(\vec{a})= \langle \vec{a}|\varrho^{}_{\text{eq}}|\vec{a}\rangle
   +  \langle -\vec{a}|\varrho^{}_{\text{eq}}|-\vec{a}\rangle .
     \end{equation}
 Formula (\ref{A:Init-non}) shows that after a non-selective measurement of this type, the
  qubit is prepared in a pure state $|\vec{b}\rangle$. Note also that the qubit and the bath
   are completely uncoupled since the bath state does not depend on the qubit
    state, and vice versa. Although the bath density matrix
    (\ref{A:B-init-1}) formally depends on the basis state $|\vec{a}\rangle$ determining the
     effects $F^{}_{i}$ in Eqs.~(\ref{Omega-non}), it is easy to see that
      $ \varrho^{}_{B}(\vec{a})$ in fact is independent of $|\vec{a}\rangle$.
      Indeed, since the states $|\vec{a}\rangle $ and $|-\vec{a}\rangle$ form
      an orthonormal basis, Eq.~(\ref{A:B-init-1}) may be rewritten as
       $  \varrho^{}_{B}(\vec{a})=\text{Tr}_{S}\varrho^{}_{\text{eq}}$.
      If so, the trace can  now be calculated with any other orthonormal basis.
       In particular, it can be done with the  canonical states (\ref{qbit-can-bas}),
        so that the initial bath density matrix takes a universal form
           \begin{equation}
              \label{A:B-init-can}
            \varrho^{}_{B}=
   \langle 0|\varrho^{}_{\text{eq}}|0\rangle
   +  \langle 1|\varrho^{}_{\text{eq}}|1\rangle.
           \end{equation}

      Let us now assume that the qubit is initially prepared in some
       pure state $|\psi\rangle$  through a \textit{selective\/} measurement.
       Then the initial state of the composite system is given by formula
        (\ref{rho-SB-selec}) or, what is the same, by
             \begin{equation}
       \label{A:Init-sel}
      \varrho^{}_{SB}(0)=
    |\psi\rangle \langle \psi|\otimes \varrho^{\prime}_{B}(\psi)
     \end{equation}
    with the initial density matrix of the bath
       \begin{equation}
          \label{A:B-init-sel}
         \varrho^{\prime}_{B}(\psi)=
         \frac{\langle\psi|\varrho^{}_{\text{eq}}|\psi\rangle}
              {\text{Tr}^{}_{B}\langle\psi|\varrho^{}_{\text{eq}}|\psi\rangle}.
       \end{equation}
 In this case the bath carries information on the qubit
state since its density matrix depends on
  $|\psi\rangle$ through the interaction term in the Hamiltonian. Consequently,
      the product (\ref{A:Init-sel}) should be interpreted as a
      \textit{correlated\/} state of the composite system~\cite{MMR2012}.
       For the dephasing model (\ref{H}), the bath density matrix (\ref{A:B-init-sel})
       can be written in a more transparent form. To do this, we note
      that the equilibrium density matrix $\varrho^{}_{\text{eq}}$ of the composite
       system is diagonal with respect to the canonical qubit states (\ref{qbit-can-bas}).
       Then, writing $|\psi\rangle $ as a decomposition
        $|\psi\rangle=c^{}_{0}|0\rangle + c^{}_{1}|1\rangle$, we obtain for the
        density matrix (\ref{A:B-init-sel}):
          \begin{equation}
              \label{A:B-sel-expl}
              \varrho^{\prime}_{B}(\psi)=
    \frac{
    |c^{}_{0}|^2  \langle 0|\varrho^{}_{\text{eq}}|0\rangle
    + |c^{}_{1}|^2  \langle 1|\varrho^{}_{\text{eq}}|1\rangle
         }
         { |c^{}_{0}|^2  \text{Tr}^{}_{B}\langle 0|\varrho^{}_{\text{eq}}|0\rangle
         +  |c^{}_{1}|^2 \text{Tr}^{}_{B}\langle 1|\varrho^{}_{\text{eq}}|1\rangle}.
          \end{equation}
   Since, in general, the bath density matrices
   (\ref{A:B-init-can}) and (\ref{A:B-init-sel})  differ from each other,
    the time evolution of the composite system depends on the type of the
     preparation measurement. Suppose, however, that the qubit
      is prepared selectively in a pure state $|\psi\rangle$ with
       equal populations of the canonical basis states, i.e., with
      $\langle\sigma^{}_{3}\rangle\equiv \langle \psi|\sigma^{}_{3}|\psi\rangle=0$.
      In this case we have $ |c^{}_{0}|^2=|c^{}_{1}|^2$, so that
       the density matrices (\ref{A:B-init-can}) and (\ref{A:B-sel-expl}) coincide.
  This is the reason why the decoherence dynamics for the non-selective
   measurement scheme (\ref{Omega-non}) is identical to the decoherence dynamics for a
    selective measurement with $\langle \sigma^{}_{3}\rangle=0$.


\begin{thebibliography}{99}
\bibitem{PRA2010}
A.~Smirne, H.-P.~Breuer, J.~Piilo, and B.~Vacchini, Phys. Rev. A
\textbf{82}, 062114 (2010).
\bibitem{PRA2012}
V.~Semin, I.~Sinayskiy and F.~Petruccione, Phys. Rev. A
\textbf{86}, 062114 (2012).
\bibitem{PRA2013}
A.~Z.~Chaudhry and J.~Gong, Phys. Rev. A \textbf{87}, 012129
(2013).
\bibitem{MMR2012} V.G.~Morozov, S.~Mathey and  G.~R\"opke, Phys. Rev. A \textbf{85}, 022101 (2012).
\bibitem{Luczka}
J.~Dajka, B.~Gardas, and J.~\L uczka, Int. J. Theor. Phys. \textbf
{52}, 1148 (2013).
\bibitem{q-measurement1}
H.M.~Wiseman, G.J.~Milburn {\it Quantum measurement and control}
(Cambridge University Press, Cambridge, 2009).
\bibitem{NL2012}
D.J.~Wineland, Rev. Mod. Phys. \textbf {85}, 1103 (2013).
\bibitem{39min} K.~Saeedi {\it et al.}, Science \textbf{342}, 830 (2013).
\bibitem{PRA2013-2}
A.~Z.~Chaudhry, and J.~Gong, Phys. Rev. A \textbf{88}, 052107
(2013).
\bibitem{BP-Book}
H.-P.~Breuer and F.~Petruccione, {\it The Theory of Open Quantum
Systems} (Oxford University, Oxford, 2002).
\bibitem{q-measurement}
V.B.~Braginsky and F.Ya.~Khalili, {\it Quantum measurement}
(Cambridge University Press, Cambridge, 1992).
\bibitem{Krauss} K.~Kraus, {\it States, Effects, and Operations}, Lecture Notes in
Physics, Vol. 190 (Springer, Berlin, 1983).
\bibitem{Holevo2001}
A.S.~Holevo, {\it Statistical structure of quantum theory},
Lecture Notes in Physics, Monographs, M67, (Springer, Berlin,
2001).
\bibitem{SRM} S.~Huang, Phys. Rev. A \textbf{72} , 022324 (2005).
\bibitem{Luczka1} J.~\L uczka, Phys. A (Amsterdam, Neth.) \textbf{167}, 919 (1990).
\bibitem{Unruh}
W.G.~Unruh, Phys. Rev. A \textbf{51}, 992 (1995).
\bibitem{MR-CMP2012} V.G.~Morozov, G.~R\"opke, Condens. Matter Phys. \textbf{15}, 43004 (2012).
\bibitem{myCMP}
V.V.~Ignatyuk, V.G.~Morozov, Condens. Matter Phys. \textbf{16}, 34001 (2013).
\bibitem{22inPRA12}
L.~Allen and J.~H. Eberly, {\it Optical Resonance and Two-Level
Atoms} (Dover, New York, 1975).
\bibitem{23inPRA12}
C. Cohen-Tannoudji, B. Diu, and F. Lalo\"e, {\it Quantum
Mechanics} (Wiley, New York, 1977).
\bibitem{Leggett}
A.J.~Leggett {\it et al.}, Rev. Mod. Phys. \textbf{59}, 1 (1987).
\bibitem{[2]inDajka}
P.~Zanardi,  D.A.~Lidar, Phys. Rev. A \textbf{70}, 012315 (2004).
\bibitem{Chaudhry14}
A.~Z.~Chaudhry and J.~Gong, Phys. Rev. A \textbf{90}, 012101 (2014).
\bibitem{2inPRA12}
U.~Weiss, {\it Quantum Dissipative Systems} (World Scientific,
Singapore, 1999).
\bibitem{14inPRA12}
H.~Grabert, P.~Schramm, and G.L.~Ingold, Phys. Rep. \textbf{168},
115 (1988).
\bibitem{18inPRA12}
L.D. Romero and J.P.~Paz, Phys. Rev. A \textbf{55}, 4070 (1997).
\bibitem{24inPRA12}
D.I.~Schuster {\it et al.}, Nature (London) \textbf{445}, 515
(2007).
\bibitem{25inPRA12}
A.D.~Cronin, J.~Schmiedmayer, and D.E.~Pritchard, Rev. Mod. Phys.
\textbf{81}, 1051 (2009).
%
\end{thebibliography}
\end{document}